\documentclass[twocolumn,showpacs,preprintnumbers,amsmath,amssymb,prl]{revtex4-1}
\usepackage{amsfonts}
\usepackage[english]{babel}
\usepackage[T1]{fontenc}
\usepackage{times}
\usepackage{mathrsfs}
\usepackage{graphicx}
\usepackage{dcolumn}
\usepackage{bm}
\usepackage[colorlinks,bookmarks=false,citecolor=blue,linkcolor=red,urlcolor=blue]{hyperref}

\def\be{\begin{equation}}
\def\ee{\end{equation}}
\def\bea{\begin{eqnarray}}
\def\eea{\end{eqnarray}}

\begin{document}

\title{Operator content of real-space entanglement spectra at conformal critical points}

\author{Andreas M. L\"auchli}
\email{andreas.laeuchli@uibk.ac.at}
\affiliation{Institut f\"ur Theoretische Physik, Universit\"at Innsbruck, Technikerstra\ss e~25, A-6020 Innsbruck, Austria}
\date{\today}

\begin{abstract}
We provide numerical evidence that the low-lying part of the entanglement spectrum of a real-space block ({\em i.e.}~a single interval) of a 
one-dimensional quantum many body system at a conformal critical point corresponds to the energy spectrum of a {\em boundary} 
conformal field theory (CFT). This correspondence allows to uncover a subset of the operator content of a conformal field theory by 
inspection of the entanglement spectrum of a single wave function, thus providing important information on a CFT beyond its central charge. 
As a practical application we show that for many systems described by a compactified boson CFT, one can infer the compactification radius (governing e.g. the power 
law decay of correlation functions) of the theory in a simple way based on the entanglement spectrum. 
\end{abstract}

\maketitle

{\em Introduction.---} 
The study of entanglement properties of quantum many body systems has established
itself as a powerful approach to uncover many new aspects of strongly correlated quantum
systems~\cite{Amico2008,Eisert2010}. 
An initial focus was on the understanding of entanglement entropies, but 
in recent years the concept of the {\em entanglement spectrum (ES)} -- introduced by Li and 
Haldane in Ref.~\cite{Li2008} --  attracted also significant interest. One of the main questions asked 
in this context is whether the spectrum of a reduced density matrix contains more 
physical information than an entanglement entropy. For fractional quantum Hall wave functions 
for example, the entanglement spectrum contains additional information on the  
theory describing the chiral edge modes, which is not accessible in entanglement entropies~\cite{Li2008}. 
In the meantime the ES has become a powerful tool to analyze topological states of matter and 
symmetry protected topological phases (such as topological insulators and Haldane insulators)
as well as non-topological states, such as e.g. continuous symmetry breaking states~\cite{ES_Work}.

One of the first areas where entanglement entropies were calculated, are systems described by 
conformal field theory, especially in 1+1 dimensions, where the conformal symmetry is particulary
rich~\cite{Belavin1984,CFTYellowBook,MussardoBook}. The properties of entanglement entropies 
in 1+1 dimensional CFTs are
quite well understood by now~\cite{Holzhey1994,Vidal2003,Korepin2004,Calabrese2004,Laflorencie2006,Cardy2010,Calabrese2010a,Calabrese2010b,Calabrese2010b,Fagotti2011}, and perhaps
best highlighted by the celebrated formula for the entanglement entropy between an interval A and its
complement in a finite system [here given for the case of periodic boundary conditions (PBC)]:
\be
S_L(L_A) \equiv -\mathrm{Tr} [\rho_A \log \rho_A] = \frac{c}{3} \log \left[ \frac{L}{\pi} \sin\left(\frac{\pi L_A}{L}\right) \right] + s_1',
\ee
where $\rho_A=\mathrm{Tr}_{L\backslash A} |\psi\rangle\langle\psi|$, $L_A$ $(L)$ is the length of the interval A 
(the entire system), and $s_1'$ is a non-universal constant. An important feature of this formula is that it only depends on the central charge $c$, and no other 
property of the CFT enters. 

\begin{figure}[t]
\includegraphics[width=0.9\linewidth]{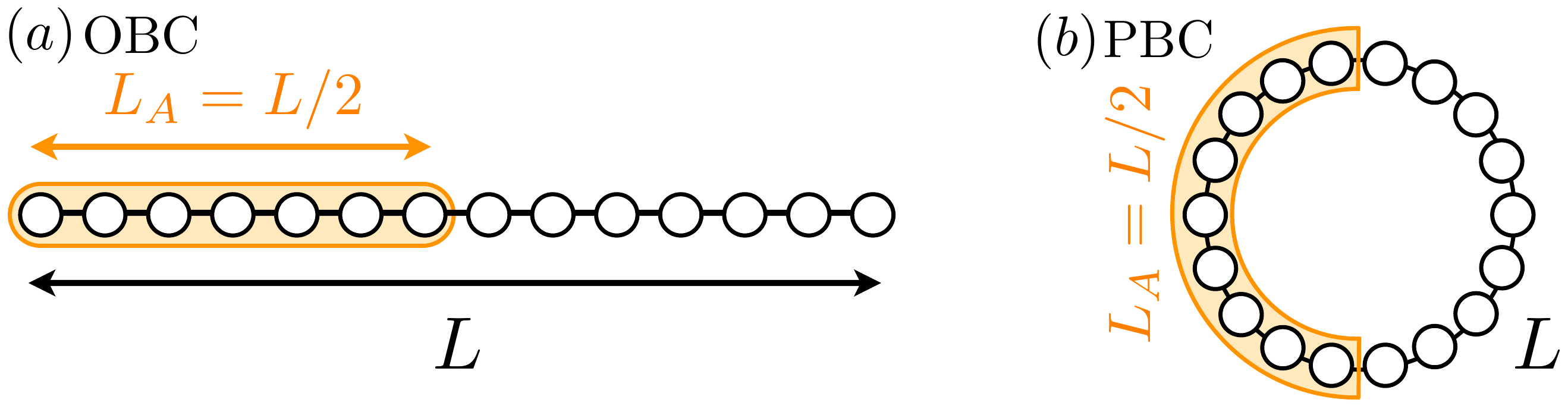}
\caption{(Color online). Setup of the bipartitions and 
 boundary conditions of the one dimensional
lattice models investigated here. The orange shaded 
region denotes the size and position of block A in the 
case of (a) open boundary conditions (OBC) and (b)
periodic boundary conditions (PBC).
} 
\label{fig:setup}
\end{figure} 
\begin{figure*}
\includegraphics[width=0.95\linewidth]{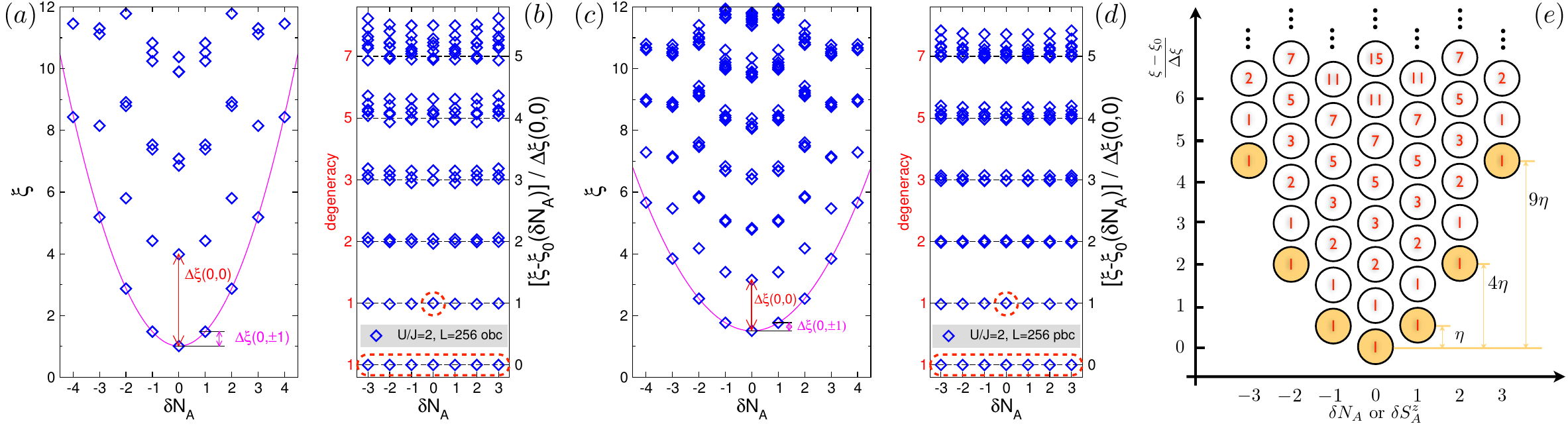}
\caption{(Color online)  
(a) / (c) Entanglement spectrum $\xi$ for a $L=256$ Bose-Hubbard chain at $U/J=2$ with OBC / PBC
and a block size of $L_A=128$. 
(b) / (d): Structure of the ES in each $\delta N_A$ sector for OBC / PBC, obtained after subtracting the 
lowest $\xi$ value in each $\delta N_A$ sector, and setting the difference between the first and second $\xi$ value in the 
$\delta N_A=0$ sector to one. The levels with an assigned value are indicated by the dashed red line, while the relative position
of all other levels highlights the emergent CFT structure of the ES. The approximate degeneracy at energy level $l$ is compatible 
with $p(l)$, the number of integer partitions of $l$. 
(e) Schematic representation of the {\em boundary} CFT energy spectrum of a compactified boson with free boundary conditions~\cite{CFTYellowBook,Alcaraz1987,Cazalilla2004}. Each orange 
shaded circle denotes a primary field with scaling dimension $\eta (\delta N_A)^2 $, while the equally spaced white circles on top 
of each primary field complete the Virasoro towers with a degeneracy count of $p(l)$. The schematic spectrum has been plotted for $\eta=1/2$. The energy differences denoted
$\Delta \xi(0,0)$ and $\Delta \xi(0,\pm1)$ in (a) and (c) enter the formula \eqref{eq:eta_estimator} $\eta= {\Delta \xi(0,\pm1)} / {\Delta \xi(0,0)}$. 
}  
\label{fig:es_bose}
\end{figure*} 
\begin{figure}[b]
\includegraphics[width=0.85\linewidth]{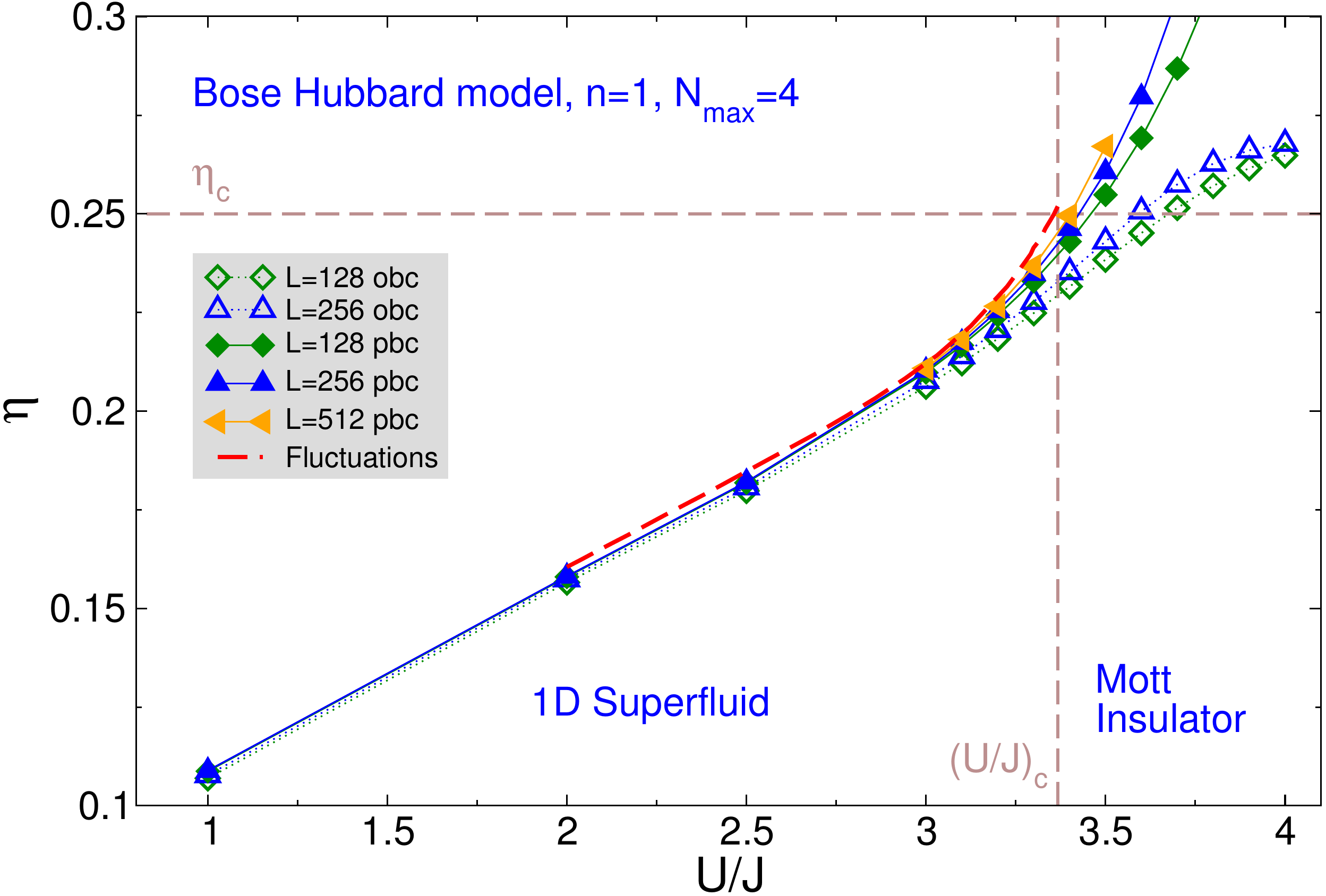}
\caption{(Color online) 
Scaling dimension $\eta$ of the Bose-Hubbard chain at filling $n=N/L=1$ as a function of 
the interaction $U/J$. The critical value $\eta_c=1/4$ and location of the transition from the superfluid to
the Mott insulator $(U/J)_c$~\cite{Kuehner1998,GiamarchiBook} are indicated with horizontal and vertical lines. The red dashed  curve
shows $\eta$ from the particle number fluctuations analysis in Ref.~\cite{Rachel2012}. The particle number
cutoff per site is set to $N_\mathrm{max}=4$ in all the results.
} 
\label{fig:Bose_eta}
\end{figure} 

In the context of 1+1D CFT it is still an open question whether the entanglement spectrum $\{\xi\}$, i.e.~minus the logarithm of
the eigenvalues of $\rho_A$: $\{\lambda\}=\{\exp(-\xi)\}$ contains additional useful physical information beyond the central
charge. Calabrese and Lef\`evre~\cite{Calabrese2008} calculated the density 
of states of the entanglement spectrum in CFT, which turned out to be universal and to depend only on the central charge. 
So the answer seems to be negative at first sight, but their results applies to the inner, continuous part of the entanglement spectrum, 
and not directly to the discrete part at the lower end of the spectrum. 

It is the aim of this paper to provide compelling numerical evidence that the lower part of real-space ES  contains interesting CFT structure by showing
several examples where the ES can be understood as the energy spectrum of a boundary CFT on a strip of effective width $\sim \log L$. Since the
energy spectra of boundary CFTs are organized in terms of conformal families~\cite{Cardy1989,Saleur1989,CardyBCFT},
one obtains access to a subset of the operator content of the underlying CFT by inspection of the ES. In the case of $c=1$ theories corresponding to a compactified boson, a recurrent task is to infer the 
compactification radius of the theory for a given microscopic model. Here we show that the boundary conditions of the boundary CFT and therefore 
the operator content of many interesting systems are such that the compactification radius (governing e.g. the exponent of correlation functions) can 
actually be deduced by a simple procedure from the ES.
 
We start by describing the simulation and entanglement spectrum setup 
and then discuss the observed operator content of the entanglement spectrum in 
various Luttinger liquids such as the Bose-Hubbard chain, the 
$S=1/2$ XXZ chain and the $S=1$ Heisenberg chain in a magnetic field,
as well as the the transverse field Ising model and the quantum three-state Potts model.
We then provide a CFT interpration of the numerical results and conclude.

{\em Setup.---} 
We study one-dimensional quantum many body systems on chains of 
length $L$ with open boundary conditions (OBC) and periodic boundary
conditions (PBC) in the following. The fixed size and position of the block
A in the bipartition is illustrated in Fig.~\ref{fig:setup}. We calculate the finite size ground state wave 
functions using the density matrix renormalization group (DMRG) technique~\cite{DMRG} and
study the entanglement spectrum for various system sizes up to about a thousand sites. The results
are converged in the number of states kept, where this number varies among the models but can
be as large as a few thousand states.

{\em Bose Hubbard model.---} We investigate the entanglement spectrum of 
Bose-Hubbard chains of length $L$ described by the Hamiltonian:
\be
H_\mathrm{BH}=- J \sum_{\langle i,j\rangle} (b_i^\dagger b_j + h.c.) + U/2 \sum_i n_i (n_i-1),
\ee
at fixed unit filling $\langle n_i\rangle=1$. Below a critical value of $U/J \approx 3.38$~\cite{Kuehner1998} the system 
is in a 1D superfluid phase, described by Luttinger liquid theory - i.e. an instance of a $c=1$ CFT -
where some of the scaling dimensions of the theory vary continuously with the ratio $U/J$~\cite{GiamarchiBook}.

In Fig.~\ref{fig:es_bose}(a) we display the ES $\{\xi\}$ for a symmetric bipartition of an OBC wave function
at $L=256$ and $U/J=2$ using the particle number in the block $(\delta N_A:=N_A-L_A)$ as an additional label of the entanglement levels.
A prominent feature of the ES is that it exhibits a parabolic envelope highlighted by the continuous line 
through the lowest ES level in each $\delta N_A$ sector. 
In order to uncover the additional structure above the parabola we subtract the value of
lowest ES in each sector, and set the overall scale by assigning the second ES level in the $\delta N_A=0$ sector a 
fixed value of one. The resulting spectrum is shown in Fig.~\ref{fig:es_bose}(b), displaying an intriguing equally spaced
structure, where furthermore the approximate degeneracy at level $l$ seems to be given by $p(l)$, the number of
integer partitions of $l$.

Before interpreting these results we show the ES for a chain of the same size and Hamiltonian parameters,
but with periodic boundary conditions in Fig.~\ref{fig:es_bose}(c) [corresponding to setup (b) in Fig.~\ref{fig:setup}]. The ES also shows
a parabolic envelope, but the spectrum is about a factor two denser compared to the OBC case.
Still, if one applies the shifting and rescaling procedure as in the OBC case one finds in Fig.~\ref{fig:es_bose}(d) the same equally spaced spectrum
with an identical degeneracy count $p(l)$ at level $l$. It is quite remarkable that the ES of the PBC setup is 
isostructural to the OBC case, because in gapped systems the structure of the entanglement spectra in the two setups is generically 
different~\cite{Alba2012a}.

In order to provide an interpretation of the structure of the entanglement spectrum it is instructive to recall the form of the
low energy spectrum of a Bose-Hubbard model in the superfluid phase on a finite system with {\em open} boundary conditions.
For example in Ref.~\cite{Cazalilla2004} the corresponding energy spectrum has been derived using bosonization techniques.
The energy spectrum has the form sketched in Fig.~\ref{fig:es_bose}(e),
which indeed matches the observed structure both in the OBC and PBC entanglement spectrum. This is one of the key results of this paper:
the entanglement spectrum has a close correspondence to the energy spectrum of a {\em boundary} CFT, but with specific - here free - boundary conditions, both
for OBC and PBC of the entire system. 

In the CFT language this spectrum contains a collection of primary fields (orange shaded circles)
which form the root of their respective Virasoro tower of descendent fields~\cite{CFTYellowBook,Alcaraz1987}. Most interestingly the relative position of the primary fields expressed in units of the
spacing of $\hat{L}_0$ (denoted $\Delta \xi$ here) encodes their nontrivial scaling dimensions. Those are
related to the Luttinger liquid parameter (or in a related way the compactification radius of the boson or the decay exponent of the Green's function).
Based on this insight, we can now devise a procedure to extract the scaling dimensions directly from the ES. If we define the 
$\hat{L}_0$-spacing in the ES by $\Delta \xi(0,0):=\xi_1(\delta N_A=0)-\xi_0(\delta N_A=0)$ [as indicated in Figs.~\ref{fig:es_bose}(a) and (c)] and the
position of the first non-identity primary field as $\Delta \xi(0,\pm1):=\xi_0(\delta N_A=\pm 1)-\xi_0(\delta N_A=0)$ [as indicated in Figs.~\ref{fig:es_bose}(a) and (c)],
then we can estimate its scaling dimension as:
\be
\eta=\frac{\Delta \xi(0,\pm1)}{\Delta \xi(0,0)}.
\label{eq:eta_estimator}
\ee
In Fig.~\ref{fig:Bose_eta} we plot $\eta$ for different system sizes and boundary conditions as a function of $U/J$. For comparison we also
plot the estimate for the same quantity obtained from a recent particle number fluctuations analysis~\cite{Rachel2012}. The agreement between the different boundary condition choices
and system sizes and the reference data is very good, particularly away from the phase transition. Upon approaching the quantum phase transition from the superfluid 
to the Mott insulator the finite size effects become more pronounced, but still the $U/J$ crossing points when $\eta$ intersects the $\eta_c=1/4$ line (the quantum phase transition is triggered according to Sine-Gordon theory when $\eta$ crosses $\eta_c=1/4$~\cite{NoteEta,GiamarchiBook,Kuehner1998}) appear to converge to the correct transition point as system sizes are increased. 


\begin{figure}[t]
\includegraphics[width=0.9\linewidth]{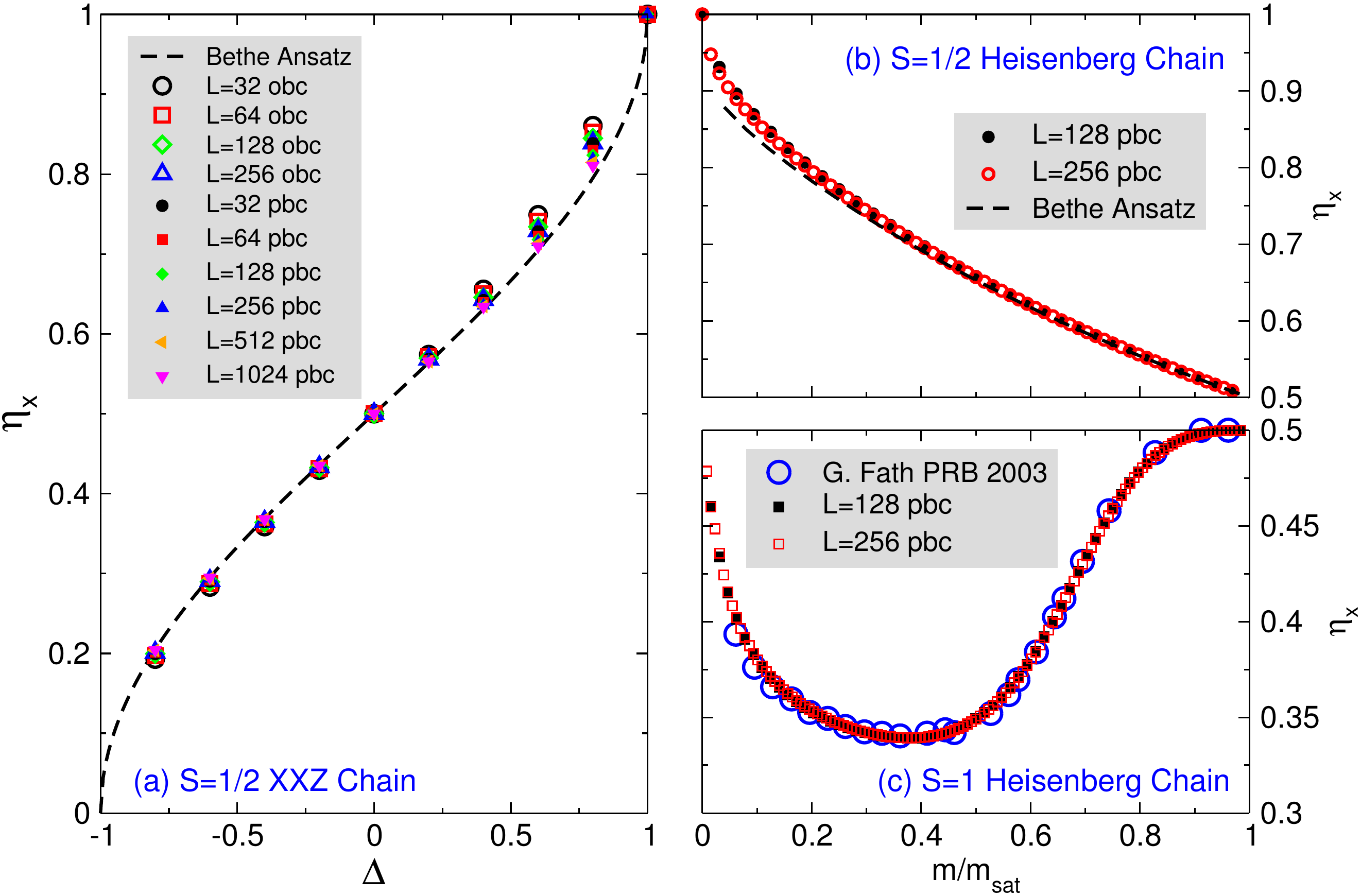}
\caption{(Color online) (a)
Scaling dimension $\eta_x$ of the $S=1/2$ XXZ model in zero magnetic field \eqref{eq:Ham_XXZ} obtained from Eq.~\eqref{eq:eta_estimator} 
as a function of $\Delta=J_z/J_{xy}$ for different system sizes and boundary conditions. The dashed line denotes the exact Bethe ansatz result
for $\eta_x$.
(b) Scaling dimension $\eta_x$ of the 
$S=1/2$ Heisenberg model (XXZ model at $\Delta=1$) as a function of the magnetization per site. 
The dashed line denotes the numerically exact Bethe ansatz result for $\eta_x$ from Ref.~\cite{Essler2004}.
(c) The same quantity $\eta_x$ of the $S=1$ Heisenberg chain as a function of the magnetization per site. 
The blue circles denote the DMRG data of Ref.~\cite{Fath2003}, where the exponent $\eta_x$ was determined based on fits of the decay of the correlation functions.
} 
\label{fig:scalingdims_spinchain}
\end{figure} 
\begin{figure*}[t]
\includegraphics[width=0.9\linewidth]{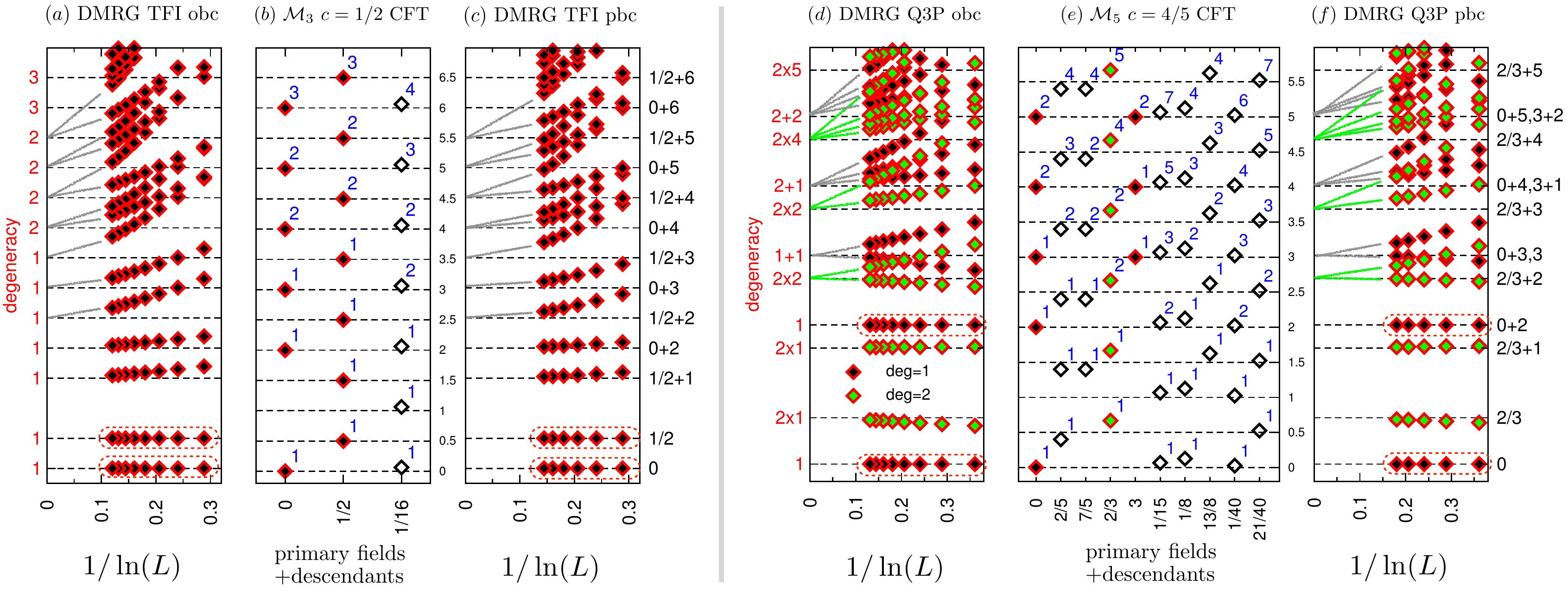}
\caption{(Color online) 
Left panel: (a) / (c) shifted and rescaled ES for the transverse field Ising model at the critical point for 
OBC / PBC. The levels which have assigned values are encircled by the dashed red line. (b) catalog 
of the conformal towers of the $\mathcal{M}_3$ $c=1/2$ CFT. The filled symbols denote the two towers
realized in the numerical ES.
Right panel: (d) / (f) shifted and rescaled ES for the quantum three-state Potts model at the critical point for 
OBC / PBC. The levels which have assigned values are encircled by the dashed red line. (e) catalog 
of the conformal towers of the $\mathcal{M}_5$ $c=4/5$ CFT. The filled symbols denote the three distinct towers
realized in the numerical ES.
}
\label{fig:tfi_q3p_es_cft}
\end{figure*} 

{\em Spin chains.---} We now investigate whether the ES based extraction of the value of $\eta$ is also feasible and accurate for quantum spin 
 systems. First we study the $S=1/2$ XXZ model:
\be
H_\mathrm{XXZ} = \sum_{\langle i,j\rangle} \frac{J_{xy}}{2}\left ( S_i^+ S_j^- +h.c. \right) +  J_{z}\ S^z_i S^z_j + h \sum_i  S^z_i
\label{eq:Ham_XXZ}
\ee
This model is integrable by Bethe Ansatz and has a Luttinger liquid ground state for $-1 < \Delta=J_z/J_{xy} \le 1$ at $h=0$. 
The dependence of the scaling dimension $\eta_x$ on $\Delta$ (governing the power-law decay of the transverse spin correlations) is known exactly from the Bethe ansatz: $\eta_x(\Delta)=1-1/\pi \arccos(\Delta)$ for $h=0$.
In Fig.~\ref{fig:scalingdims_spinchain}(a) we show the exact curve together with the DMRG ES results for different system sizes and boundary 
conditions. Similar to the Bose-Hubbard case the agreement is excellent for small $\eta_x$ and also improves systematically with increasing system size. 
We  tentatively attribute the discrepancy between the ES estimate for $\eta_x$ and the exact results at larger $\Delta$ to more pronounced finite size effects in the ES due to an operator becoming marginal 
as $\Delta \rightarrow 1$, similar to the Bose-Hubbard chain at the transition to the Mott insulator. Let us note that the scheme \eqref{eq:eta_estimator}
reproduces the exact result for $\eta_x$ already for finite system size both at $\Delta=0$ ($\eta_x=1/2$) due to the exactly known entanglement spectrum in terms of free
fermions~\cite{Peschel2004,Peschel2009}, and at $\Delta=1$ ($\eta_x=1$), due to the multiplet structure of the $SU(2)$ symmetry which is also present in the entanglement
spectrum. 

As a further application of the method we show the magnetization dependence of $\eta_x$ for the Heisenberg model ($\Delta=1$) in a magnetic field in Fig.~\ref{fig:scalingdims_spinchain}(b).
For comparison the numerical Bethe ansatz results from Ref.~\cite{Essler2004} are shown. 
Finally in Fig.~\ref{fig:scalingdims_spinchain}(c) we present the magnetization dependence of
$\eta_x$ in the $S=1$ Heisenberg chain as a function of the magnetization, where an interesting non-monotonous behavior  of $\eta_x$ has been reported in 
past work~\cite{Sakai1991,CamposVenuti2002,Fath2003,Friedrich2007}. We also include the results obtained in Ref.~\cite{Fath2003} based on fits to the correlation functions
for comparison, highlighting the good agreement between the two complementary approaches. We believe though that our ES based method is simpler to 
use than earlier numerical results based on correlation function fits. 

In order to ascertain that the correspondence between the ES and a boundary CFT extends beyond Luttinger liquids with microscopic particle number 
conservation, we investigate the transverse field Ising and the quantum three-state Potts chain as further examples.

{\em The transverse field Ising model.---} The transverse field Ising chain at its
critical point: :
\be
H_\mathrm{TFI}= - \sum_i \sigma^z_i \sigma^z_{i+1} + \sum_i \sigma^x,
\ee
realizes the Ising CFT with central charge $c=1/2$. DMRG results for the entanglement
spectrum for both OBC (with free BC) and PBC are
shown in Fig.~\ref{fig:tfi_q3p_es_cft}(a) and (c) respectively. We shift
the lowest $\xi$ level to zero and subsequently rescale the entire spectrum such that the second 
level is set to $1/2$. The resulting ES is plotted as a function of $1/\ln(L)$.
The rescaling is motivated by the structure which becomes apparent
in the limit $1/\ln L \rightarrow 0$. The numerical data then plausibly suggests that the ES arranges on a grid
of integer and half integer values, with an approximate degeneracy count which
can be interpreted as the sum of the two Virasoro towers $0\oplus 1/2$, see Fig.~\ref{fig:tfi_q3p_es_cft}(b)
for a table of the operator content of the Ising CFT, where the realized fields are represented by 
filled symbols. It is remarkable that this identification matches the operator content of the partition function $Z_{ff}$ 
of the Ising CFT on an annulus with free boundary conditions on both sides~\cite{Cardy1989,CardyBCFT}. 
We also stress that - as in the compactified boson cases studied before - the ES for both 
OBC and PBC feature the same operator content. By modifying the boundary conditions in the OBC setup 
(e.g. by applying a magnetic field in the $z$ direction at both ends) it is possible to change the operator content 
and reveal a single tower with scaling dimension $1/16$~\cite{Bonnes2013}, in agreement with~\cite{Cardy1989}.

Finally we note that one can also calculate the ES using the mapping to free fermions~\cite{Chung2001,Peschel2009}, 
that a relation between Virasoro characters and the spectrum of reduced density matrices based on the corner transfer
matrix off criticality has been discussed previously~\cite{CardyLesHouches,Lepori2013} and in 
Ref.~\cite{Peschel1987} the spectrum of a critical corner transfer matrix was studied in a different setup, but a counting 
similar to ours was observed.

{\em The quantum three-state Potts model.---} As a last example we study the
quantum three-state Potts model on a chain at its critical point~\cite{QuantumPotts}. This lattice model realizes a CFT 
with central charge $c=4/5$. We plot the ES data obtained by DMRG using OBC (with free BC) and PBC based on a
similar rescaling procedure, but this time we choose to scale the fourth level to a fixed value of two. The resulting ES
spectra are displayed in Fig.~\ref{fig:tfi_q3p_es_cft}(d) and (f). This rescaling is justified a posteriori by the
emergent structure. Comparing the catalog of primary fields and their descendents of the $\mathcal{M}_5$
minimal model CFT~\cite{CFTYellowBook,MussardoBook} [shown in Fig.~\ref{fig:tfi_q3p_es_cft}(e)] with the ES 
data it is plausible that the ES has the operator content: $0\oplus (2\times 2/3) \oplus 3$. This
operator content is identical to the one of the CFT annulus partition function $Z_{ff}$ of the Potts model with free
boundary conditions~\cite{Saleur1989,Cardy1989}. By modifying the boundary conditions in the OBC setup 
(e.g. by applying a local field preferring or suppressing one of the three states at both ends) it is possible to change 
the operator content and reveal the towers $1/8\oplus13/8$ or $1/40\oplus21/40$~\cite{Bonnes2013}, in agreement 
with~\cite{Saleur1989,Cardy1989}.

{\em CFT interpretation.---} We have provided substantial numerical evidence that 
the entanglement spectrum of quantum lattice models at a conformal critical point can be interpreted as the spectrum of the $\hat{L}_0$ operator in a {\em boundary} CFT. The idea
that the ES of a region is related to $\hat{L}_0$ appears repeatedly in the  CFT entanglement literature~\cite{Holzhey1994,Orus2005,Orus2006,DeChiara2012,Lepori2013},
but the specific boundary conditions and the resulting operator content have not been clearly stated. In the models studied in the present
work we find the conformal boundary conditions to be free. The identical operator content of the OBC and PBC setup 
provides further evidence that for these systems the boundary condition at the entanglement cut is a free boundary condition. Free boundaries
at the entanglement cut also appear naturally in the replica approach by Calabrese and Cardy~\cite{Calabrese2004,PasqualePrivate}.
An analytical derivation for the numerical observations reported here is needed, and might also address the
origin and possible additional useful information encoded in the finite-size corrections to the ES (such as marginal operators). It will 
also be important to understand whether different microscopic implementations of the same CFT can lead to different boundary conditions
in the boundary CFT and thus modify the operator content of the ES. 

In this paper we have shown that the real space entanglement spectrum gives access to the
operator content (or equivalently the energy spectrum) of certain boundary CFTs. In the particular case of those
Luttinger liquids considered in this work it is possible to extract the Luttinger liquid parameter solely based on the 
entanglement spectrum. This technique has the potential to become a standard diagnostic tool in DMRG or related
matrix product state based techniques (such as iTEBD~\cite{iTEBD}, iDMRG~\cite{iDMRG} or cMPS~\cite{cMPS}, possibly with different CFT 
boundary conditions and operator content~\cite{Bonnes2013}) to characterize one-dimensional conformally invariant systems or 
multicomponent generalizations thereof (such as fermionic Hubbard models with gapless spin and charge degrees of freedom).

%

\acknowledgments

I acknowledge an inspiring discussion with M. Metlitski which motivated the present work.
I am very grateful to J.~Dubail for discussions on several aspects of CFT and for emphasizing the
importance of the boundary conditions. I thank  L.~Bonnes, P.~Calabrese, T.~Grover, G.~Mussardo, I.~Peschel, A.~Turner and F.~Verstraete 
for discussions and V.~Alba, L.~Bonnes and M.~Haque for related work.
The DMRG simulations have been performed on machines of the platform "Scientific computing" at the University of
Innsbruck - supported by the BMWF. I acknowledge support through the Austrian Science Foundation (FWF) SFB Focus (F40-18).


\begin{thebibliography}{99}

\bibitem{Amico2008}
L. Amico, R. Fazio, A. Osterloh, and V. Vedral, 
Rev. Mod. Phys. {\bf 80}, 517 (2008).

\bibitem{Eisert2010}
J. Eisert, M. Cramer, and M. B. Plenio
Rev. Mod. Phys. {\bf 82}, 277 (2010).

\bibitem{Li2008}
H. Li and F.D.M. Haldane,
Phys. Rev. Lett. {\bf 101}, 010504 (2008).

\bibitem{ES_Work}
N.~Regnault, B.~A.~Bernevig, F.~D.~M.~ Haldane, Phys.\ Rev.\ Lett. {\bf 103}, 016801 (2009);
%
N.~Bray-Ali, L.~Ding, and S.~Haas, Phys.\ Rev.\ B {\bf 80}, 180504(R) (2009);
%
L.~Fidkowski, Phys.\ Rev.\ Lett.\ {\bf 104}, 130502 (2010);
%
A.~M.~L\"auchli, E.~J.~Bergholtz, J.~Suorsa, and M.~Haque, Phys.\ Rev.\ Lett.\ {\bf 104}, 156404 (2010);
%
R.~Thomale, A.~Sterdyniak, N.~Regnault, and B.~A.~Bernevig, Phys.\ Rev.\ Lett.\ {\bf 104}, 180502 (2010);
%
H.~Yao and X.~L.~Qi, Phys.\ Rev.\ Lett.\ {\bf 105}, 080501 (2010);
%
E.~Prodan, T.~L.~Hughes, and B.~A.~Bernevig, Phys.\ Rev.\ Lett.\ {\bf 105}, 115501 (2010);
%
F.~Pollmann, A.~M.~Turner, E.~Berg, M.~Oshikawa, Phys.\ Rev.\ B, {\bf 81}, 064439 (2010);
%
M.~Kargarian and G.~A.~Fiete, Phys.\ Rev.\ B, {\bf 82}, 085106 (2010);
%
A.~M.~Turner, Y.~Zhang, A.~Vishwanath, Phys.\ Rev.\ B, {\bf 82}, 241102R (2010);
%
Z.~Papic, B.~A.~Bernevig, and N.~Regnault, Phys.\ Rev.\ Lett.\ {\bf 106}, 056801 (2011);
%
L.~Fidkowski, T.~S.~Jackson and I.~Klich, Phys.\ Rev.\ Lett.\ {\bf 107}, 036601 (2011);
%
J.~Dubail, and N.~Read, Phys.\ Rev.\ Lett.\ {\bf 107}, 157001 (2011);
%
J.~Schliemann, Phys.\ Rev.\ B {\bf 83}, 115322 (2011);
%
J.~I.~Cirac, D.~Poilblanc, N.~Schuch, and F.~Verstraete, Phys.\ Rev.\ B {\bf 83}, 245134 (2011);
%
T.~L.~Hughes, E.~Prodan, B.~A.~Bernevig, Phys.\ Rev.\ B, {\bf 83}, 245132 (2011);
%
N.~Regnault and B.~A.~Bernevig, Phys.\ Rev.\ X {\bf 1}, 021014 (2011);
%
M.~A.~Metlitski and T.~Grover, arXiv:1112.5166;
%
X.~L.~Qi, H.~Katsura, and A.~W.~W.~Ludwig, Phys.\ Rev.\ Lett.\ {\bf 108}, 196402 (2012);
%
D. Poilblanc, N. Schuch, D. Perez-Garcia, and J.I. Cirac, Phys. Rev. B {\bf 86}, 014404 (2012);
%
J.~Dubail, N.~Read, and E.H. Rezayi,
Phys. Rev. B {\bf 85}, 115321 (2012);
%
A. Sterdyniak, A. Chandran, N. Regnault, B. A. Bernevig, and P. Bonderson,
Phys. Rev. B {\bf 85}, 125308 (2012);
%
I.D.~Rodriguez, S.H. Simon, and J. K. Slingerland, 
Phys. Rev. Lett. {\bf 108}, 256806 (2012).
%
N. Schuch, D. Poilblanc, J.~I.~Cirac, and D.~Perez-Garcia, arXiv:1210.5601 (unpublished);
%
V. Alba, M. Haque and A.M. L\"auchli,
arxiv:1212.5634 (unpublished).

\bibitem{Belavin1984}
A.A.~Belavin, A.M.~Polyakov and A.B.~Zamolodchikov,
Nucl. Phys. B {\bf 241}, 333 (1984).

\bibitem{CFTYellowBook}
P.~di Francesco, P.~Mathieu, and D.~S\'en\'echal,
{\em Conformal Field Theory}, 
(Springer, 1997).

\bibitem{MussardoBook}
G.~Mussardo,
{\em Statistical Field Theory},
Oxford Graduate Texts, Oxford University Press (2009)

\bibitem{Holzhey1994}
C. Holzhey, F. Larsen, and F. Wilczek, 
Nucl. Phys. B {\bf 424}, 443 (1994).

\bibitem{Vidal2003}
G.~Vidal, J.I.~Latorre, E.~Rico, and A.~Kitaev,
Phys. Rev. Lett. {\bf 90}, 227902 (2003).

\bibitem{Korepin2004}
V.E.~Korepin,
Phys. Rev. Lett. {\bf 92}, 096402 (2004).

\bibitem{Calabrese2004}
P.~Calabrese, and J.~Cardy,
J. Stat. Mech. P06002, (2004).

\bibitem{Laflorencie2006}
N.~Laflorencie, E.S.~S{\o}rensen, M.-S.~Chang, and I.~Affleck,
Phys. Rev. Lett. {\bf 96}, 100603 (2006).

\bibitem{Cardy2010}
J.~Cardy and P.~Calabrese,
J. Stat. Mech. P04023 (2010).

\bibitem{Calabrese2010a}
P.~Calabrese and F.H.L.~Essler,
J. Stat. Mech. P08029 (2010).

\bibitem{Calabrese2010b}
P.~Calabrese, M.~Campostrini, F.~Essler, and B.~Nienhuis,
Phys. Rev. Lett. {\bf 104}, 095701 (2010).

\bibitem{Fagotti2011}
M.~Fagotti and P.~Calabrese,
J. Stat. Mech. P01017 (2011).

\bibitem{Calabrese2008}
P. Calabrese and A. Lef\`evre,
Phys. Rev. A {\bf 78}, 032329 (2008).


\bibitem{Cardy1989}
J. Cardy, 
Nucl. Phys. B {\bf 324}, 581, (1989).

\bibitem{Saleur1989}
H. Saleur and M. Bauer,
Nucl. Phys. B {\bf 320}, 591 (1989).

\bibitem{CardyBCFT}
J. Cardy,
{\em Boundary Conformal Field Theory},
in "Encyclopedia of Mathematical Physics", J.-P. Fran\c{c}oise, G. Naber and T.S. Tsun, eds. (Elsevier, 2005.)
(arXiv:hep-th/0411189)

\bibitem{DMRG}
S.R. White, 
Phys. Rev. Lett. {\bf 69}, 2863 (1992);
U. Schollw\"ock, 
Rev. Mod. Phys. {\bf 77}, 259 (2005).

\bibitem{Kuehner1998}
T.D.~K\"uhner and H.~Monien,
Phys. Rev. B {\bf 58}, R14741 (1998).

\bibitem{GiamarchiBook}
T.~Giamarchi,
{\em Quantum physics in one dimensions},
Oxford university press, Oxford (2004). 

\bibitem{Alba2012a}
V.~Alba, M.~Haque, and A.M. L\"auchli, 
Phys. Rev. Lett. {\bf 108}, 227201 (2012).

\bibitem{Cazalilla2004}
M.A.~Cazalilla,
Journal of Physics B: Atomic, Molecular and Optical Physics {\bf 37}, S1 (2004).

\bibitem{Alcaraz1987}
F.C. Alcaraz, M.N. Barber, M.T. Batchelor, R.J. Baxter and G.R.W Quispel,
J. Phys. A: Math. Gen. {\bf 20}, 6397 (1987).

\bibitem{Rachel2012}
S. Rachel, N. Laflorencie, H.F. Song, and K. Le Hur,
Phys. Rev. Lett. {\bf 108}, 116401 (2012).

\bibitem{NoteEta}
For the Bose-Hubbard model this quantity $\eta$ is identical to the decay exponent of the bosonic Green's function $\langle b^\dagger_i b_{i+r}\rangle~\sim~|r|^{-\eta}$.

\bibitem{Peschel2004}
I.~Peschel,
J. Stat. Mech. P06004 (2004).

\bibitem{Peschel2009}
I.~Peschel, and V.~Eisler,
J. Phys. A: Math. Theor. {\bf 42}, 504003 (2009).

\bibitem{Essler2004}
F.H.L. Essler and R.M. Konik,
arXiv:cond-mat/0412421, published in the I. Kogan Memorial Volume by World Scientific.

\bibitem{Sakai1991}
T.~Sakai and M.~Takahashi, 
Phys. Rev. B {\bf 43}, 13383 (1991).

\bibitem{CamposVenuti2002}
L.~Campos Venuti, E.~Ercolessi, G.~Morandi, P.~Pieri, and M.~Roncaglia,
Int. J. Mod. Phys. B Vol. {\bf 16}, 1363 (2002).

\bibitem{Fath2003}
G. F\'ath,
Phys. Rev. B {\bf 68}, 134445 (2003).

\bibitem{Friedrich2007}
A.~Friedrich, A.K.~Kolezhuk, I.P.~McCulloch, and U.~Schollw\"ock,
Phys. Rev. B {\bf 75}, 094414 (2007).

\bibitem{Bonnes2013}
L.~Bonnes and A.M. L\"auchli, unpublished.

\bibitem{Chung2001}
M.-C.~Chung and I.~Peschel,
Phys. Rev. B {\bf 64}, 064412 (2001).

\bibitem{CardyLesHouches}
J. Cardy, 
{\em Conformal Invariance and Statistical Mechanics}, 
Lecture Notes in "Fields, Strings and Critical Behavior, proceedings of les Houches Summer School in Theoretical Physics, 1988"
(North-Holland, 1990).

\bibitem{Lepori2013}
L.~Lepori, G.~De Chiara, and A.~Sanpera,
arXiv:1302.5285, unpublished

\bibitem{Peschel1987}
I.~Peschel and T.T.~Troung,
Z. Phys. B {\bf 69}, 395 (1987).

\bibitem{QuantumPotts}
The microscopic Hamiltonian is as in:
A.~Rapp, P.~Schmitteckert, G.~Takacs, G.~Zarand,
New J. of Phys. {\bf 15}, 013058 (2013).

\bibitem{Orus2005}
R.~Or\'us,
Phys. Rev. A {\bf 71}, 052327 (2005).

\bibitem{Orus2006}
R.~Or\'us, J.I.~Latorre, J.~Eisert, and M.~Cramer,
Phys. Rev. A {\bf 73}, 060303 (2006).

\bibitem{DeChiara2012}
G. De Chiara, L. Lepori, M. Lewenstein, and A. Sanpera,
Phys. Rev. Lett. {\bf 109}, 237208 (2012).

\bibitem{PasqualePrivate}
P. Calabrese, private communication.

\bibitem{iTEBD}
G.~Vidal,
Phys. Rev. Lett. {\bf 98}, 070201 (2007).

\bibitem{iDMRG}
I. P. McCulloch,
arXiv:0804.2509, unpublished

\bibitem{cMPS}
F. Verstraete and J. I. Cirac,
Phys. Rev. Lett. {\bf 104}, 190405 (2010).


%
%

\end{thebibliography}
\end{document}